\documentclass[sigconf]{acmart}
\usepackage{graphicx}
\usepackage{tabularx}
\usepackage{multirow}

\AtBeginDocument{%
  }

\copyrightyear{2024}
\acmYear{2024}
\setcopyright{licensedothergov}\acmConference[ICMLC 2024]{2024 16th
International Conference on Machine Learning and Computing}{February 2--5,
2024}{Shenzhen, China}
\acmBooktitle{2024 16th International Conference on Machine Learning and
Computing (ICMLC 2024), February 2--5, 2024, Shenzhen, China}
\acmDOI{10.1145/3651671.3651693}
\acmISBN{979-8-4007-0923-4/24/02}

\begin{document}

\title{Temporal Convolution-based Hybrid Model Approach with Representation Learning for Real-Time Acoustic Anomaly Detection}


\author{Sahan Dissanayaka}
\orcid{0009-0004-0972-6200}
\affiliation{%
  \institution{University of Colombo School of Computing}
  \streetaddress{}
  \city{Colombo}
  \country{Sri Lanka}}
\email{tsahandisaai@gmail.com}

\author{Manjusri Wickramasinghe}
\orcid{0000-0002-0725-5124}
\affiliation{%
  \institution{University of Colombo School of Computing}
  \streetaddress{}
  \city{Colombo}
  \country{Sri Lanka}}
\email{mie@ucsc.cmb.ac.lk}

\author{Pasindu Marasinghe}
\orcid{0000-0001-6350-3391}
\affiliation{%
  \institution{University of Colombo School of Computing}
  \streetaddress{}
  \city{Colombo}
  \country{Sri Lanka}}
\email{ppm@ucsc.cmb.ac.lk}

\renewcommand{\shortauthors}{Dissanayaka et al.}

\begin{abstract}
  The early detection of potential failures in industrial machinery components is paramount for ensuring the reliability and safety of operations, thereby preserving Machine Condition Monitoring (MCM). This research addresses this imperative by introducing an innovative approach to Real-Time Acoustic Anomaly Detection. Our method combines semi-supervised temporal convolution with representation learning and a hybrid model strategy with Temporal Convolutional Networks (TCN) to handle various intricate anomaly patterns found in acoustic data effectively. The proposed model demonstrates superior performance compared to established research in the field, underscoring the effectiveness of this approach. Not only do we present quantitative evidence of its superiority, but we also employ visual representations, such as t-SNE plots, to further substantiate the model's efficacy.
\end{abstract}



\received{30 September 2023}
\received[revised]{07 November 2023}
\received[accepted]{10 September 2023}

\maketitle

\section{Introduction}
Analysis of acoustic data is vital because it carries a large amount of information about our everyday environment and the physical events that take place. Concerning research, acoustic-related studies aligned with a focus on the development of high-level physical or computational models to infer properties of the environments \cite{ruff2021}. This trend has seen significant growth in recent times, with the improvement of deep learning models and high commercial valuation \cite{ane2021}. Such applications span various domains, encompassing anomaly detection, event detection, event localization, scene classification, sound generation, etc. However, concerning acoustic anomaly detection, extensive research is still required to consistently detect and recognize acoustic characteristics close to human auditory perception \cite{spiousas2017}. Human hearing capability remains superior in the presence of several overlapping noises, detecting rare anomalies present in simultaneous and distorted surroundings. An intelligent solution that copes with deep learning hybrid models has a huge potential to reach this level, whereas it has the capability of tackling different yet complex data patterns. This study focuses on improving the capability of computers for MCM in industrial systems to recognize anomaly activities in their environments using acoustic information as near to the human expert level using such hybrid deep neural model architectures.

\subsection{Anomaly Detection}
Anomaly detection is a technique that identifies the data which deviates from its normal behaviour. Anomaly points allude to the observations which deviate from its normal behaviour. The importance of anomaly detection comes that, an anomaly can indicate important events, such as production faults, delivery bottlenecks, system defects, or heart flicker, and is therefore of central interest. The above high-level understanding can be formally represented with the support of the probability theorem. It needs two main aspects of anomaly detection: First, the "concept of normality behaviour" and Second, "deviations or exceptions of the data." 

\begin{center}
\begin{figure}[htbp]
\centerline{\includegraphics[scale=0.3]{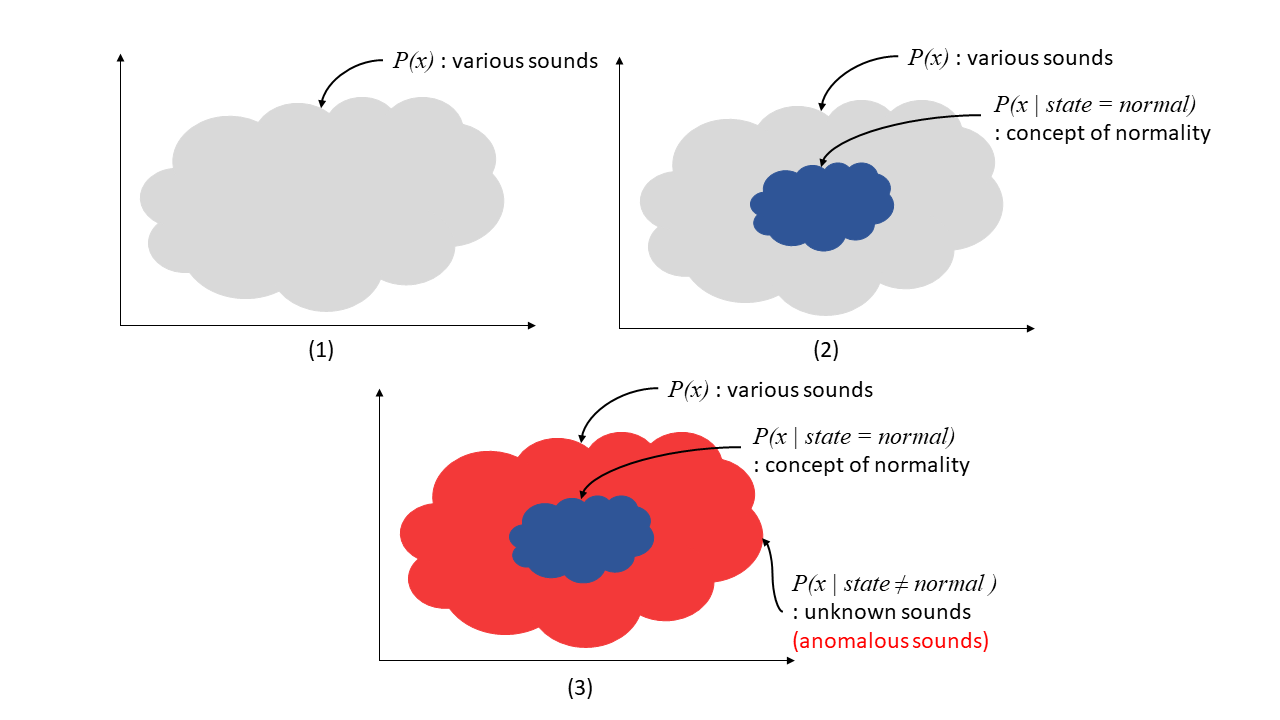}}
\caption{Defining high-level intuition of anomaly detection into a formalized model using probability theory concepts mentioned in Equation 1.1 with support of two aspects which are "concept of normality" and "deviations or exception of the data ."(1) Defines the hypothetical data space which contains both normal and anomaly instance. Outside the context(cloud) depicted the area which is irrelevant to the considered domain. (2) Defines the identified normal data distribution and the rest of space inherently becomes the region of unknown/not accepted with normal(anomaly) data region which is denoted in red on (3). The threshold level(size of the blue region) is highly domain-specific and needs delicate work to determine it.}
\label{fig:anomaly-cloud}
\vspace{-0.5em}
\end{figure}
\end{center}

So, let us define the data space by $X$, where $ X \subseteq \mathbb{R}^D $ and the concept of normality as a distribution of $\mathbb{P}^+$ on $X$. If the data instance $x$ has a deviation or exception from the law of normality, it lies on a low probability region in a standard distribution. By assuming $\mathbb{P}^+$ has a probability density function of $p^+(x)$, anomalies (A) can be denoted by with the threshold of $\tau$ as follows,
\begin{equation}
    A = \{x \subseteq X   |   p^+(x) \leq \tau \} , \tau \geq 0
\end{equation}
The threshold level manages the region of anomalies sufficiently into a small space in order to create the boundary between normal and anomaly instances \cite{ruff2021}. Also, that value is highly application-specific. Proper validation can verify the effectiveness of the defined threshold value at the application level. Observe Figure \ref{fig:anomaly-cloud} for clarity.

\vspace{-1.5em}

\subsection{Anomaly Types on Acoustic Data}
Several studies have proposed different taxonomies for anomalies as a convention for anomaly detection. The initial model was proposed by Chandola et al. \cite{chandola2009} with three forms, namely \textbf{point}, \textbf{contextual} and \textbf{collective} anomalies. When it comes to low-level and high-level sensory anomalies populated with sensors, Ruff et al. \cite{ruff2021} add on two other categories of anomalies as \textbf{Sensor anomalies} and \textbf{Semantic anomalies}. This study follows three types of anomaly convention proposed by Chandola et al. \cite{chandola2009} because it is the widely acknowledged model for anomaly detection studies. Plus, it is well aligned with the related studies under acoustic anomaly detection, and this study does not use sensor-related information to tackle anomalies. An accomplished intelligent solution should ultimately address all the unique anomaly data types to give a generalizable solution.

\subsection{Real-Time Acoustic Anomaly Detection(RTAAD)}
Compared to other data types, acoustic anomalies are complex and challenging due to its multidimensionality and temporal variations; addressing them in real-time is far more challenging with respect to offline mode. Usually, such research problem domain is known as RTAAD. Further, detecting anomalies in the real-time context has a vast commercial valuation including MCM. It drops the risk of unprotected data representing important, critical, actionable information for future perspectives. Also, it anticipates(prevents) unplanned downtime, optimizes operating efficiency, and early detection of novelty behaviours for quick diagnosis. A real-world example of RTAAD is when an expert car technician is able to identify an engine fault by its anomalous sound. That observation shows that the human ear is well-trained to identify the anomaly behaviours of a sound(a waveform) with the proper expertise over time. Let's take another example to prove the limitation of human auditory perception, whereas an expert system must address some occasions than humans. When identifying the fault of a train engine while travelling, the engine driver could not hear those anomalies due to the limitation of human auditory space. However, experienced passengers who lie between the auditory space can predict the anomaly behaviour. These scenarios convey the message that acoustic anomalies have their own natures and sometimes human intervention cannot capture them all. Our motivation is to bring those human behaviours into digitized form and to use that expertise via a sophisticated Real-Time Anomaly Detection mechanism for acoustic(sound) data. The challenge here is how to deliver proactive measurements for detecting anomalies. But it is achievable by a better learning model and a suitable learning technique \cite{fpp3,aggarwal2016}. 

\subsection{Sequence Modelling with Temporal Convolution}
The reason for flowing with sequence modelling in this study is that it is well aligned with the intuition of audio data. Audio is a time series that needs to be addressed by sequence modelling \cite{Sutskever2014}. It allows the analysis of rapid change and long dependencies that must help tackle three types of anomalies proposed by \cite{chandola2009}. A Sequence model is a function $f$ that maps vector $T+1$ to another vector of $T+1$ as follows.
\begin{equation}
\hat{y}_{0}, \hat{y}_{1}, .... \hat{y}_{T-1}, \hat{y}_{T} = f(\hat{x}_{0}, \hat{x}_{1}, .... \hat{x}_{T-1}, \hat{x}_{T})
\end{equation}
Sequence modelling with temporal convolution has the advantage over typical standard convolution. It only computes the value based on past and present, and computationally it is less expensive. To elaborate more, standard convolution is costly ($O(n^2)$) and uses both future and past values to predict the present value.
\begin{equation}
 y(i, j) = \sum_{m=1}^{m}\sum_{n=1}^{n} x(i+m, j+n).w(m,n)
\end{equation}
However, temporal convolution only relies upon past values to determine the present value.
\begin{equation}
 y(t) = \sum_{m=1}^{k} x(t-k).w(k)
\end{equation}

\subsection{Reconstruction-based Anomaly Detection Models}
Typically, reconstruction-based deep learning models employ only normal data and go through an encoding procedure \cite{pang2021}. The goal is to transform the incoming data into a lower-dimensional representation before attempting to recreate it via a latent space representation. It implements through a bottleneck of the encoding and decoding process with the technique of \textbf{dimensionality reduction}. Because of that, these models shine brightest when the input data is high-dimensional and complicated, and a lower-dimensional representation would aid in capturing essential characteristics of the data. Capturing of anomaly happens when the models are not possible to reconstruct the anomaly samples since it is only trained with the normal(semi-supervised) samples. It results in a larger reconstruction error for the anomalies. It matches elegantly with the modelling requirement of our hypothesis boundary differentiation as stated in Figure \ref{fig:anomaly-cloud}.

\subsection{Representation Learning}
Also, Representation Learning(RL) is another emerging technique of learning a sophisticated distribution of normal/anomaly data. It strives to learn and eventually represent the normal boundary and comment about the anomalies in advance \cite{bengio2014representation}. Also, a good representation summarises the explanatory distribution of the data and enables the extraction of useful information while interpreting the unique properties of the entire distribution; it can be considered as a supervised predictor. RL model heavily manifested to identify the boundary of normal visually and the underlying data distribution as stated in the \ref{fig:anomaly-cloud}; hence, it was selected for further exercise during the study with Semi-Supervised Learning and Representation Learning.

Thus far, an expanding number of machine learning and deep learning algorithms have been developed to detect acoustic anomalies, but researchers merely focus on detecting acoustic anomalies in real-time \cite{pang2021, chalapathy2019}. With respect to that, the core contribution of this research is given by delivering an improved representation-based deep hybrid model architecture using a temporal convolution-based deep neural network to capture complex time series features for RTAAD and enable it to be used in MCM scenarios effectively. Further, the study attempts to give a detailed comparison between the proposed hybrid model performance with current state-of-the-art models available in the literature. Moreover, the proposed study analyzes single model performance versus performance gain achieved through the coherent hybrid model architecture. In accordance, section 2 represents our methodology follows to building the hybrid model.

\section{Methodology}
\begin{center}
\begin{figure}[htbp]
\centerline{\includegraphics[scale=0.4]{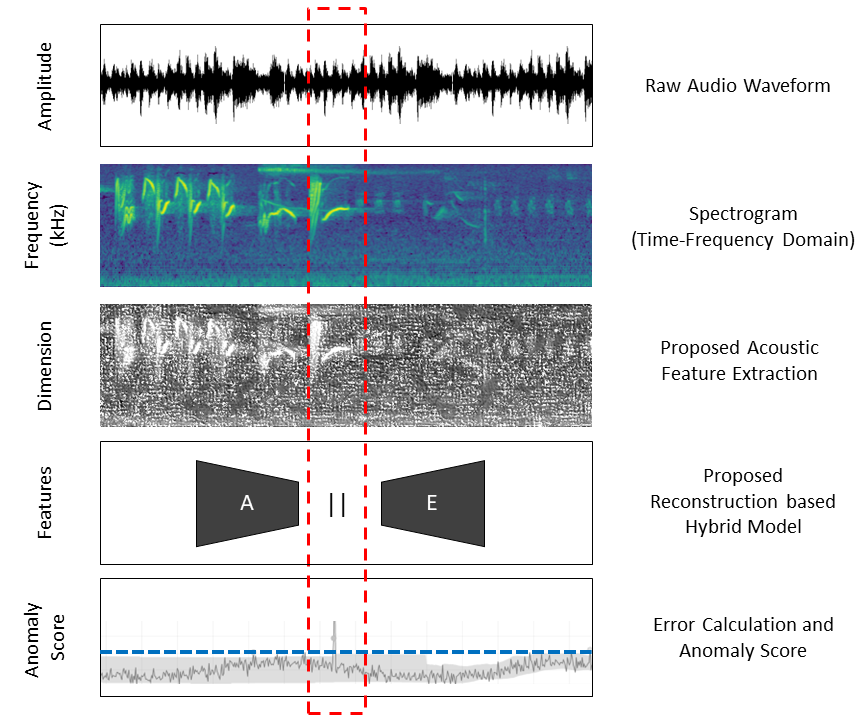}}
\caption{Layer breakdown of data in each design stage of Real-Time Acoustic Anomaly Detection problem using the proposed Deep Hybrid Model}
\label{fig:design-stages-rtaad}
\vspace{-0.5em}
\end{figure}
\end{center}

The proposed methodology aims to enhance modeling through a hybrid architecture, adept at capturing diverse anomaly patterns. It integrates suitable feature extraction methods to extract unique feature sets, addressing the challenges associated with anomaly detection. On that venture, a robust pipeline is proposed (see Figure \ref{fig:design-stages-rtaad}), drawing inspiration from Koizumi et al.'s work \cite{ntt_technical_review_2017}. This design facilitates the learning of audio and the capture of anomalies through a set of components, enhancing various stages of the anomaly detection process. The pipeline delineates the layer breakdown, illustrating how data is represented at each stage and the procedures applied from high-level acoustic waveforms to low-level features.

First, the observed raw audio waveform converts into a spectrogram account for both time and frequency domain signal features. Secondly, it goes through the feature extraction step to identify prominent characteristics in the desired audio. Usually, this step requires filter banks. This study proposed that Mel Filter Banks undergo the operation at this stage. Then, Third, extracted feature vectors feed into the proposed reconstruction-based hybrid model. The model learns the complex pattern that encompasses the vectors which pass through to the last layer which is calculating the error(reconstruction) and calculating the anomaly score of the signal. With a defined threshold level for the anomalies, it identifies and provides a way of demystifying the anomalies. The following sections will sequentially cover the procedures of each layer.

\subsection{Data Integration}
\begin{center}
\begin{figure}[htbp]
\centerline{\includegraphics[scale=0.3]{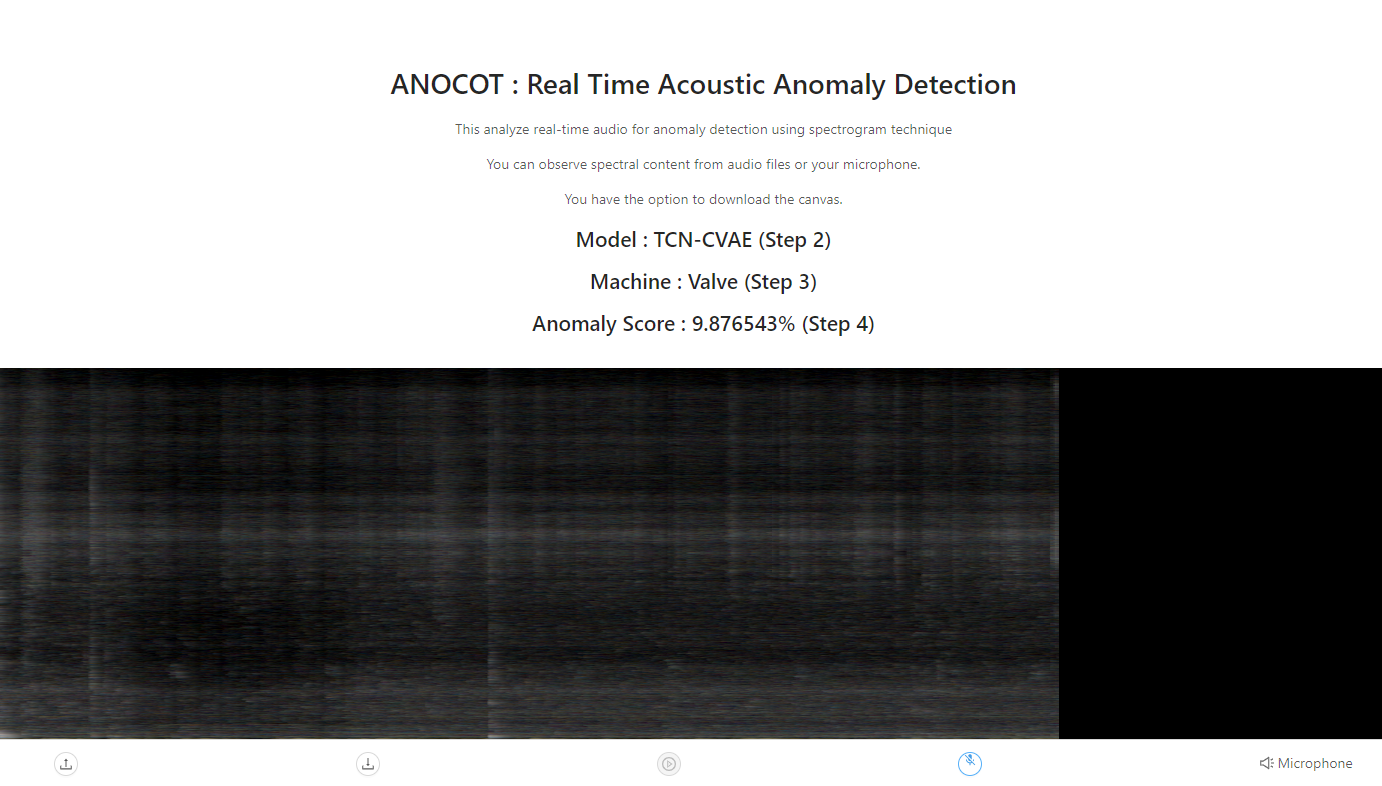}}
\caption{Implemented Real-Time Inference user interface mentioned in the below diagram. It creates a Mel-Spectrogram in real-time with a sliding windowing technique and feeds into the Python backend.}
\label{fig:rtaad-anocot-interface}
\vspace{-0.5em}
\end{figure}
\end{center}

A meticulous data collection phase precedes the model-building process to ensure optimal performance. The MIMII dataset \cite{mimii} is integrated for offline training, and for real-time inference, a raw audio signal detection mechanism is implemented, utilizing an Internal Apple Microphone with noise cancellation configurations. The real-time inference is visualized through a user interface (UI), depicted in Figure \ref{fig:rtaad-anocot-interface}.

\subsection{Feature Extraction}
A feature extraction pipeline is developed to input diverse characteristics into the hybrid model architecture. The techniques presented address common signal processing challenges, optimizing feature capture in the time-frequency domain, minimizing spectral leakage, and capturing anomalies at frame boundaries with windowing. Feature vectors are then fed into the modelling phase for data pattern analysis, with the Log-Mel spectrogram specifically incorporated for identifying unique features.

\textbf{Log-Mel Spectrogram:} A spectrogram is a visual representation of the frequency content of an audio signal over time which is similar to how humans perceive frequency logarithmically. This can be used to identify changes in the spectral content of the signal and extract features such as spectral density, onset time, and duration. It converts raw audio waveform using Short Time Fourier Transform to convert into the frequency domain and then integrates mel-filter banks to mix auditory perception. So Mel Spectrograms are, $Mel-Spectrograms = (Time-Frequency + Perceptually-Relevant)$. It uses 512 Mel Bins with 1024 n\_fft with a hop length of 512. This study integrates Librosa \cite{librosa} for all feature implementations. The conversion from spectrogram to mel scale $m$ is happened as per the following equation,

\begin{equation}
    m = 2595.log(1 + f/500)
\end{equation}

To convert frequencies into mel-scale, we define Mel-filter banks or Mel-Bins. The selection of the number of Mel-Bins is application-specific, serving as a hyper-parameter. The Experiment Section provides our rationale for choosing the number of Mel-Bins.

\subsection{Modelling Normal with Semi-Supervised Learning}
Anomalies are rare and complex instances which deviate from its normal behaviour. Attempting to identify all anomaly behaviours is not possible because anomalies can change in a completely different way than we assumed to be. Because of that, direct supervision of anomalies is not practical. On the other hand, total unsupervised doesn't guarantee of capturing all anomalies since it tries to learn the intrinsic properties and patterns present in data by itself \cite{chalapathy2019}. It may be correct or incorrect. So there is an uncertainty.

Alternatively, we can consider exclusively inputting normal instances into the model and querying the model about both normality and abnormality. Following our anomaly definition in Equation \textbf{1}, the core hypothesis of anomaly detection aligns with this concept, illustrated in Figure \ref{fig:anomaly-cloud} Part 3. This approach recognizes that focusing solely on the concept of normality facilitates assessing the distribution of normal behaviour and anomalies (area beyond the boundary of normal).

\textbf{\centerline{"Learn what normal is, then capture the anomalies"}}

It means that if our model is good enough to learn the normal boundary, it is easy for the model to realize what an anomaly is. Because anomaly patterns are not learned/seen by the model at the training stage, despite that, it performs poorly at the inference for those points. Owing to the above fact, we strive to learn \textbf{concept of normality training only with normal samples} other than learning both anomaly and normal. It saves the overhead of gathering every corner of anomalies of machine sounds. Since we are exactly sure that the model learns only with the normal instance and infers both the normal and anomaly, it follows a semi-supervised nature of learning. But the key challenge is that learn a sophisticated and accurate representation or solve the question of \textbf{"how the anomaly and normal representation reside in the data space?"} in a highly sampled audio dataset. It can be achieved through Representation Learning. Formally, representation is a visual characteristic of data that can entangle and conceal various explanatory factors or variations behind the data. It may be a straight line denoted by $Ax1 + By1 + C = 0$, a circle, or some other specialized mathematical configuration. Extracted feature vectors, as described in the last section, or any interpretable vector, help to learn the representation of the data by the deep learning model. That is how the reconstruction-based models come into the picture with dimensionality reduction techniques like t-SNE 
\cite{vanDerMaaten2008,velliangiri2019} to simplify learning and visualizing the representation.

\begin{equation}
Anomaly Score(X_{test}) = \frac{1}{p} \sum_{i=1}^{p} ||x - \hat{x_i}||^2
\end{equation}

Since both signals are extracted using spectrogram techniques above equation could be expanded as follows. It is the same equation mentioned in section 1.2. 

\begin{equation}
Anomaly Score (X\textsubscript{test}) = 1/N\textsubscript{s}\sum_{i=1}^N (X^a\textsubscript{i} - X^r\textsubscript{i})^T(X^a\textsubscript{i} - X^r\textsubscript{i})     
\end{equation}

\subsection{Implementation of TCN-VCAE Hybrid Architecture}
Hybrid models carry the fundamental idea of making a combination that compensates for the limitations of one approach with the strengths of the other. These models took the recent attention of scholars in anomaly detection paradigm with the promising results shown in the domain of time series forecasting \cite{makridakis2018, makridakis2020}. \cite{pang2021} pointed out that the applicability of ensembled deep learning models for anomaly detection received drastic performance change with respect to a single model. The Auto Encoder (AE) is perhaps the most well-known simplest model in deep learning that relies on the process of reconstruction. Going with that, our study selected AE as the baseline neural architecture for modelling on the grounds of its promising results achieved in the literature \cite{chandola2009, chalapathy2019, hojjati2022} and to well position the proposed hybrid architecture in the class reconstruction-based model class. On top of it, researchers combine the below candidate neural architectures to embed wide coverage of anomaly detection for the proposed architecture.

\textbf{Variational Auto Encoders :} The capability of detecting reduced features of the data space. It is good at finding point and contextual types of anomalies \cite{ane2021}.

\textbf{Convolutional Networks :} Good at detecting complex time series patterns but not good at detecting arbitrary long sequences.

\textbf{Recurrent Networks :} Able to handle arbitrary long sequences via recurrence but suffering from vanishing/ exploding gradient problems.

Owing to above facts, our work on hybrid model creation has two parts. The first part is the creation of hybrid-reconstruction model with AE. The second part is to add hierarchical sequential modelling capability with Temporal Convolutional Networks (TCN). In the first part, researchers started with a simple Convolutional Auto Encoder (CAE). To avoid the typical limitations of AE as it doesn't regularize the input, In the second phase, researchers add VAE as training is regularized to minimize overfitting and ensure that the latent space has desirable properties that enable the generative process as it becomes Convolutional Variational Auto Encoder (VCAE).

\subsubsection{Why TCN type of design for RTAAD Problem?}
\begin{center}
\begin{figure}[htbp]
\centerline{\includegraphics[scale=0.25]{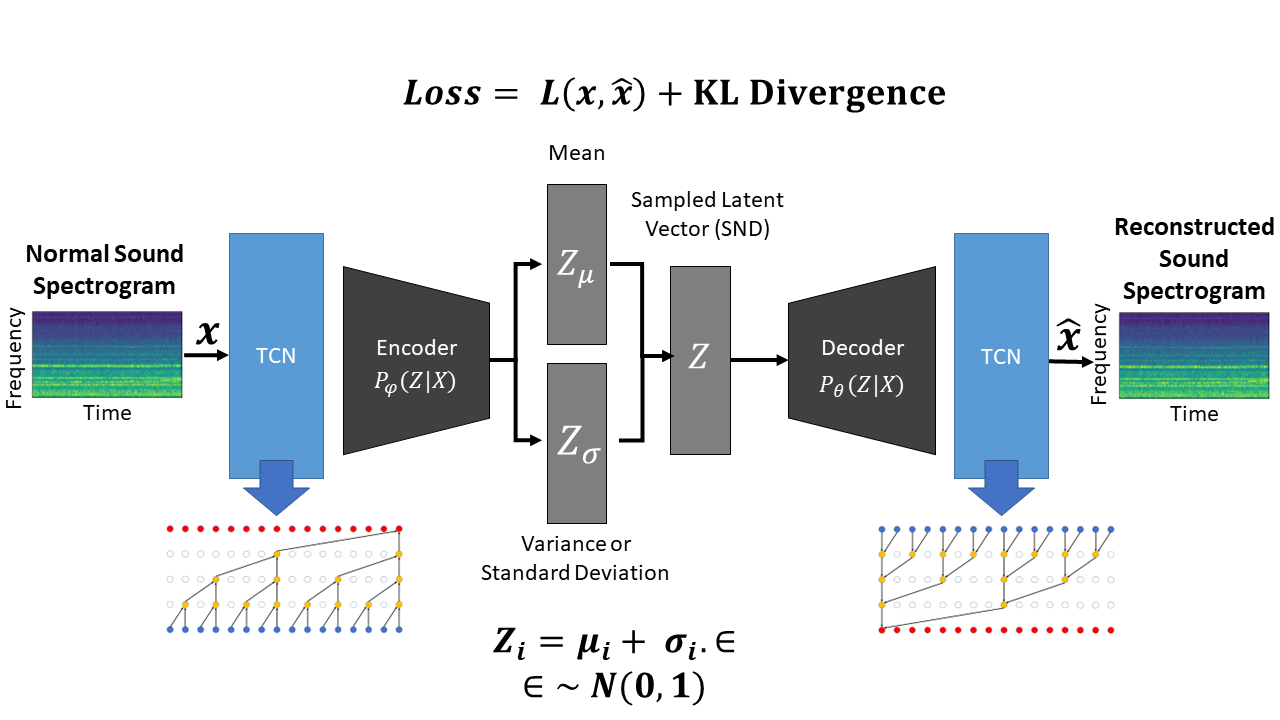}}
\caption{TCN Architectures Overview: it is never the less a dilated hierarchical convolution layer network which provides recurrent neural capability with lesser parameters}
\label{fig:tcn-arch}
\vspace{-2.5em}
\end{figure}
\end{center}

TCN \cite{BaiTCN2018} architectural nature helps in several ways to outperform recurrent neural networks in handling time series data. TCN offers a more efficient training process, requiring only one forward and one back-propagation cycle, in contrast to the resource-intensive training of RNNs. The hierarchical structure of TCN alters the gradient flow direction, resembling a quasi-recurrent neural network with dilated convolution, eliminating the need for backpropagation across time and enabling parallel training. The exponentially growing receptive field of TCN allows it to observe all input values throughout the entire time series, effectively addressing unique patterns in long dependencies, a challenge typically encountered by Recurrent Neural Networks. Furthermore, TCN avoids the exploding/vanishing gradient problem along the time axis. Figure \ref{fig:tcn-arch} illustrates the proposed architecture and essential enhancements made to the error term in this study.

In TCN network architecture, the receptive field $R$ or history coverage may be represented as follows, with the number of layers l and a convolutional kernel size of k: 

\begin{equation}
    R = 2^{l}(k-1)
\end{equation}

\centerline{$Loss_{VAE}$ = Reconstruction Loss(MSE) + KL Divergence}

\begin{equation}
Loss_{VAE}(x,\hat{x},\mu,\log\sigma) = \frac{1}{2} \sum_{i=1}^{d}(x_i - \hat{x_i})^2 + \frac{1}{2} \sum_{i=1}^{d} (1 + \log(\sigma_i^2) - \mu_i^2 - \sigma_i^2)
\end{equation}

Since the core objective of the proposed hybrid model is to tackle the three main types of anomalies that exist in the literature, whereas the proposed hybrid model comprises sub-architectures to model those three distinct anomalies in order to capture the aforementioned anomalous natures, our hybrid design technically covers the aspects of all anomaly data patterns.

\subsection{Anomaly Score Calculation}
Instead of relying on a simple "normal" or "abnormal" score (0 or 1), it is better to use a more nuanced "anomaly" value because the binary class outcomes can contain a higher error percentage. Owing to the above reason, one approach to finding anomalies in data is to compute an anomaly score. The goal is to assign a value to each data point that represents how unusual it is in relation to the rest of the dataset. If the score is high, the data point in question will likely be an outlier. There are several different approaches to calculating anomaly scores \cite{ane2021}, each approach optimized for a particular context or class of anomalies. This study involves distance-based anomaly score calculation to analyze the score. So, it is basically the difference between the inputted signal and the reconstructed signal outcome. The research shows that the reconstruction error rate via computing the parameters of the latent space (feature vector) and the reconstructed output.

\begin{center}
\begin{figure}[htbp]
\vspace{-1.5em}
\centerline{\includegraphics[scale=0.5]{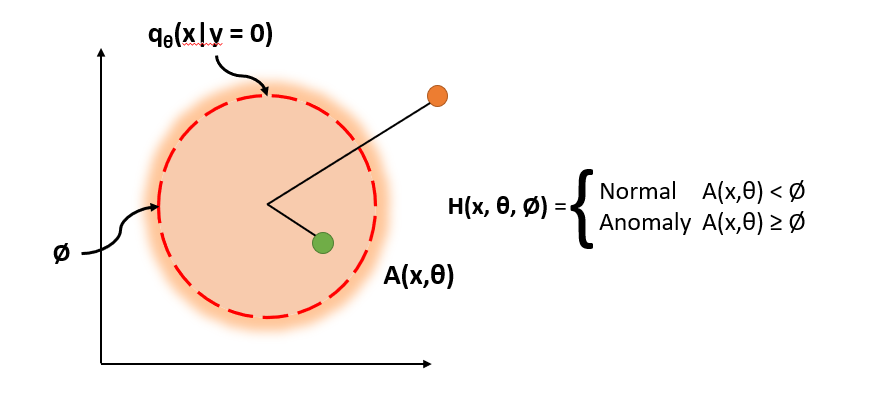}}
\caption{Anomaly Score Calculation with respect to the Hypothesis of Modelling Normal of a Data Distribution}
\label{fig:anomaly-score-calculation}
\vspace{-2.0em}
\end{figure}
\end{center}

As shown in Figure \ref{fig:anomaly-score-calculation} the anomaly score is high when the input signal displays unusual behaviour, and low otherwise. Because our initial hypothesis is AE based model that cannot reconstruct an anomaly sample because it only trains with normal samples. An anomaly score calculator $A(x)$ with a parameter in order to compute the anomaly score. $A(x)$ takes as input the operational sound $X_{test}$ of a target machine, together with information about that machine, such as its Machine Type and Machine ID, and returns a single anomalous score for the whole audio clip, denoted by the notation $A(x_{test})$. Then, x is considered anomalous if and only if the anomaly score is greater than the defined threshold. The threshold value determined by the maximum False Positive Rate(FPR) is acceptable by the model. If the researchers tolerate only 10\%, we consider only that range of values. That defines by the evaluation metric. As a result, $A$ must be trained so that $A(x)$ takes a significant value not only when the entire audio-clip x is anomalous but also when a part of x is abnormal, as with collision anomalous noises. The evaluation section details the technique used to determine the threshold. As cited in Figure \ref{fig:anomaly-score-calculation} shows how A(x) and Decision Threshold determine the anomaly and normal boundary.

\section{Experiments}

\subsection{Environment}
The Apple M1 Pro Silicon Chip was used for all the tests; it has 16 GB of RAM, an 8-core CPU, and a 6-core GPU, and it has built-in MacOS hardware for accelerating the TensorFlow GPU. All the codes are written in Python and run under the Anaconda Python distribution. Each experiment is conducted in a series fashion (minimum of three times) to reduce the likelihood of GPU mistakes. The results are averaged based on the number of times the experiment was performed.

\subsection{Dataset}
All of the project-related experiments were based on the MIMII dataset \cite{mimii}, which specifically focused on the typical(normal) and abnormal operation noises of four different types of genuine industrial machines: fan, pump, slider, and valve. Each recording is a single-channel audio file with a sampling rate ($S_r$) of 22.05 kHz and an approximate duration of ten seconds. It comprises both the operational sound of a target machine as well as noise from the surrounding environment. Moreover, there are at least seven Machine IDs and one thousand normal samples associated with each Machine ID in each machine. Additionally, at least one hundred instances of the anomaly for each Machine ID.

\subsection{Neural Network Implementation}
Keras was chosen for this research for Neural Network Implementation with Python because of its widespread community support and high computational efficiency. With respect to the TCN implementation, we adopted the initial paper working of \cite{BaiTCN2018} to implement the hybrid architecture. Also, we integrated Adam \cite{adam} optimizer for stochastic optimization of the neural networks. 

\subsection{Analyze Data Distributions via t-SNE Visualizations without Modelling}
When it comes to problems with anomaly detection, it is indispensable to observe the difference between the anomaly data space and the normal data space. This is done so that one can determine the level of deep experimentation in terms of dimensions that are necessary to differentiate the class boundaries and what the characteristic behaviours are that a particular class demonstrates as a whole.

This experiment is broken down into two distinct phases. First, view the data in a low-dimensional space (2D) only after conducting the feature extraction(without modelling); then in the second phase, it compares the exact representation after applying the proposed hybrid model simultaneously evaluating how well each represents the normal and abnormal classifications.

\vspace{-1.5em}
\begin{center}
\begin{figure}[htbp]
\centerline{\includegraphics[scale=0.40]{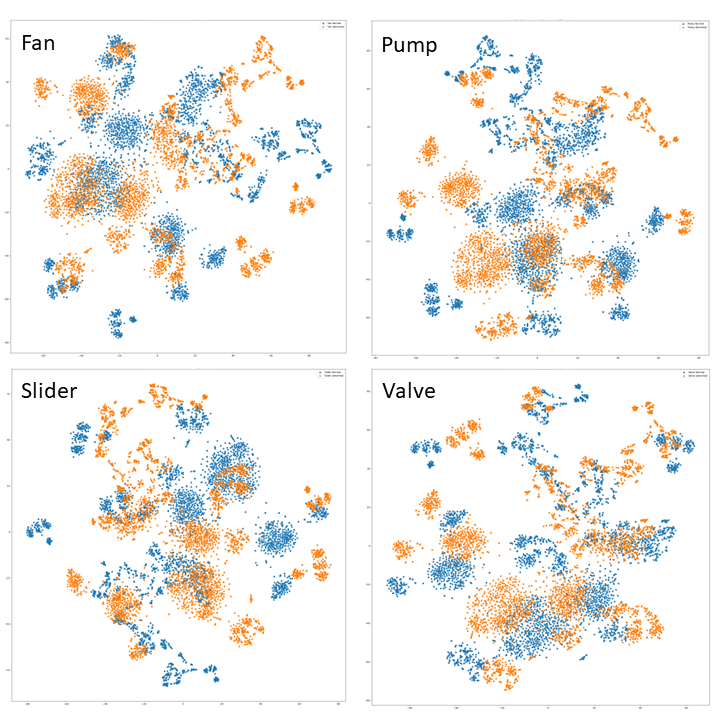}}
\caption{2D plots of four machine types of MIMII using t-SNE visualization after only applying the feature extraction. Blue data items are normal data instances and orange data items are anomaly data instances. \emph{Note: Each dot represents one data sample}}
\label{fig:tSNEmimii}
\vspace{-1.5em}
\end{figure}
\end{center}

The reader is able to recognize, as seen in the 2D plot of the t-SNE visualizations in Figure \ref{fig:tSNEmimii} that the normal areas (blue) and the anomalous areas (orange) overlapped one another in some areas. If this is the case, it is not possible to distinguish certain abnormal behaviours from normal when seen from a low-dimensional perspective. In addition, it depicts distributions of machines that do not exhibit symmetrical behaviour. The fact that this is the case raises the question of what those gaps truly signify. This initial experiment makes researchers realize the point of further suggesting the requirement for a standardized latent space model, such as the Variational Auto Encoder (VAE) irrespective of other reconstruction-based models. To get to the conclusion, "Is this tendency going to continue in higher dimensions?", we carry out another sub-stage of the experiment using 3D plot visualizations utilizing t-SNE visualizations. However, sufficient overlapping is present. It propagates the message that representation has plenty of room for improvement and modelling is necessary to achieve it.

\begin{center}
\begin{figure}[htbp]
\vspace{-0.5em}
\centerline{\includegraphics[scale=0.50]{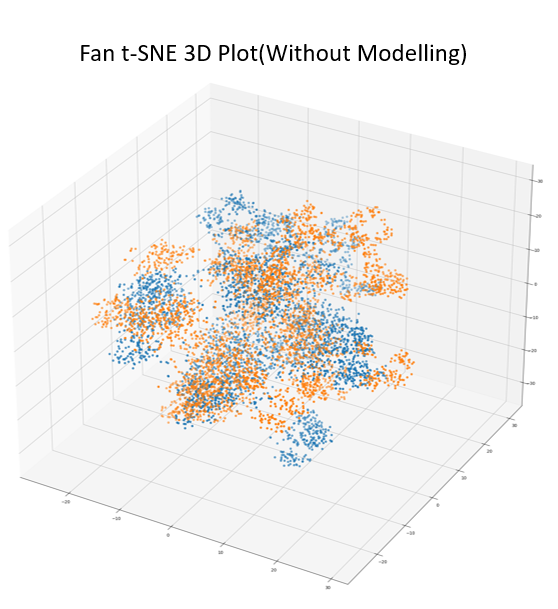}}
\caption{3D plot of Fan machine type of MIMII using t-SNE visualization after only applying the feature extraction. Blue data items are normal data instances and orange data items are anomaly data instances. \emph{Note: Each dot represents one data sample}}
\label{fig:tSNEmimii-3d-fan}
\vspace{-2.5em}
\end{figure}
\end{center}

\subsection{Improved DCASE2020 Baseline Model Results}

\vspace{-1.5em}
\begin{center}
\begin{figure}[htbp]
\centerline{\includegraphics[scale=0.25]{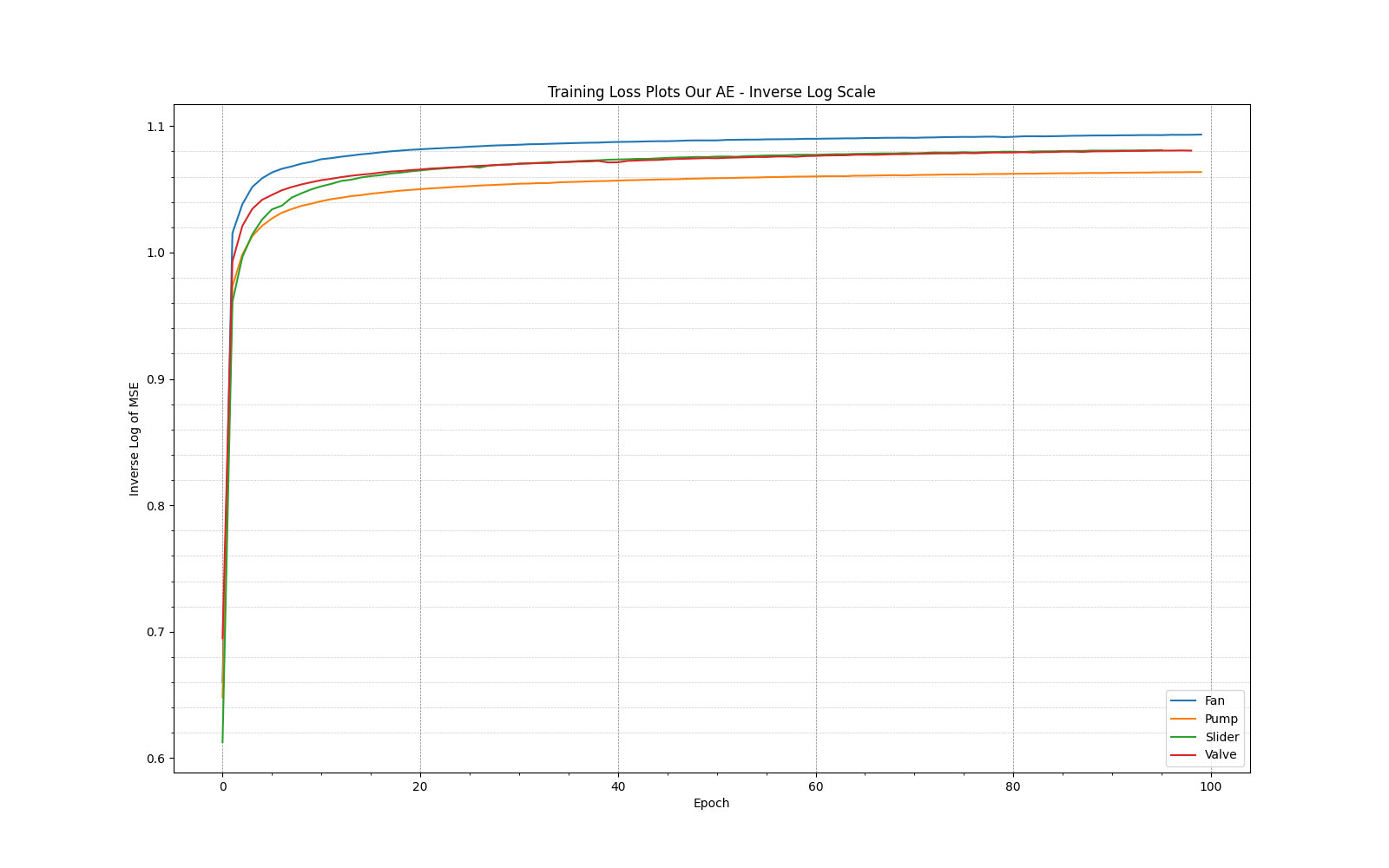}}
\caption{Our Improved Dense AE Architecture Inverse Log Scale Training Loss of All Machines}
\label{fig:all-baseline-ae-training}
\vspace{-1.5em}
\end{figure}
\end{center}

As stated in Figure \ref{fig:all-baseline-ae-training} and with respective model performance depicted in Table \ref{tab:performance-comparsion-hybrid} concluded that, compared to DCASE 2020 Baseline Training and Reconstruction results, our baseline model reduces 1.23\% training loss while increasing the reconstruction capability from 12\% to 16.5\% which is around \~4.5\% increase of reconstruction capability of normal space with respect to Valve machine type. That is a significant increase in terms of a reconstruction model. Further, it implies learning an improved representation of normal behaviour through this improvement. Even though this stage of model building is a sufficient contribution to the MIMII-related studies since our work outperforms the work of \cite{Jalali2020} and  AE model of Ribeiro et al. \cite{Ribeiro2020}. It further realizes our successfulness of robust feature extraction pipeline and modelling.

\subsection{Experiment 4: Hybrid Models Results}

Since our suggested hybrid model comprises several modelling stages (CAE to VCAE to TCN-CVAE) to improve the representation capability as well as learn the normal boundary more efficiently, we performed experiments and analyzed the data to evaluate the successful operation of each stage in the development of the hybrid model. This was done so that the inclusion of parameters and the performance of each stage could be compared. In addition, Table \ref{tab:performance-comparsion-hybrid} is given the option to track the value contributed by each model component, as well as how the model's capability to detect anomalies improves when more model strategies are combined.

{
\renewcommand{\arraystretch}{0.9}
\begin{table*}[t]
\centering
\begin{tabularx}{\linewidth}{>{\bfseries}c@{\hspace{1cm}} c@{\hspace{1cm}} c@{\hspace{0.28cm}} c@{\hspace{0.28cm}} c@{\hspace{0.28cm}} c@{\hspace{0.28cm}} c}
\hline
 Model & ID & Dense AE & Our Dense AE & CAE & CVAE & TCN-CVAE \\
\hline
Parameters & & 269,992 & 2,265,888 & 433,041 & 707,505 & 1,764,449\\
\hline
Metric &  & AUC(\%) pAUC(\%) & AUC(\%) pAUC(\%) & AUC(\%) pAUC(\%) & AUC(\%) pAUC(\%) & AUC(\%) pAUC(\%)\\
\hline
\multirow{4}{*}{Fan}   
& 0 & 54.41 \quad 49.37 & 56.73 \quad 49.72 & 51.57 \quad 49.38 &  54.22 \quad \textbf{51.76}  &  \textbf{58.64} \quad 50.17\\
& 2 & 73.40 \quad 54.81 & 79.60 \quad 53.85 & 64.41 \quad \textbf{54.09} &  63.47 \quad 52.00  &  \textbf{78.24} \quad 53.22\\
& 4 & 61.61 \quad 53.26 & 70.07 \quad 54.11 & 54.60 \quad 53.53 &  52.51 \quad 52.85  &  \textbf{70.25} \quad \textbf{54.13}\\
& 6 & 73.92 \quad 52.35 & 81.67 \quad 55.15 & 69.08 \quad \textbf{54.79} &  71.34 \quad 52.85  &  \textbf{74.02} \quad 53.70\\
& Avg & 65.83 \quad 52.45 & 72.02 \quad 53.24 &  59.92 \quad \textbf{52.95} &  59.63 \quad 51.62  &  \textbf{70.29} \quad 52.81\\
\hline
\multirow{4}{*}{Pump}  
& 0 & 67.15 \quad56.74 & 66.94 \quad 56.87 & 62.73 \quad 51.29 & 62.79 \quad \textbf{58.66} & \textbf{68.19} \quad 56.95\\
& 2 & 61.53 \quad58.10 & 60.22 \quad 60.34 & 57.68 \quad 55.46 & 56.54 \quad 54.10 &  \textbf{66.13} \quad \textbf{61.66}\\
& 4 & 88.83 \quad67.10 & 86.95 \quad 66.28 & 89.05 \quad 75.00 & 87.37 \quad 73.42 &  \textbf{93.08} \quad \textbf{76.32}\\
& 6 & 74.55 \quad58.02 & 77.53 \quad 60.32 & 64.07 \quad 52.69 & 64.00 \quad 52.06 &  \textbf{72.83} \quad \textbf{55.37}\\
& Avg & 72.89 \quad 59.99 & 72.91 \quad 60.94 & 68.39 \quad 58.60 & 67.68 \quad 58.06 &  \textbf{75.06} \quad \textbf{62.58}\\
\hline
\multirow{4}{*}{Slider}
& 0 & 96.19 \quad 81.44 & 96.12 \quad 82.30 & 90.49 \quad 69.24 & 95.43 \quad 77.42 &  \textbf{96.30} \quad \textbf{81.90}\\
& 2 & 78.97 \quad 63.68 & 79.55 \quad 64.42 & 74.25 \quad 56.54 & 78.59 \quad 57.04 &  \textbf{80.23} \quad \textbf{60.95}\\
& 4 & 94.30 \quad 71.98 & 95.44 \quad 76.14 & 90.48 \quad 72.08 & \textbf{95.76} \quad \textbf{78.06} &  95.66 \quad 77.23\\
& 6 & 69.59 \quad 49.02 & 77.22 \quad 49.52 & 69.65 \quad 50.64 & 70.02 \quad \textbf{51.67} &  \textbf{78.71} \quad 51.03\\
& Avg & 84.76 \quad 66.53 & 87.08 \quad 68.09 & 81.22 \quad 62.12 & 84.95 \quad 65.55 &  \textbf{87.73} \quad \textbf{67.78}\\
\hline
\multirow{4}{*}{Valve} 
& 0 & 68.76 \quad 51.70 & 74.61 \quad 52.28 & 76.14 \quad \textbf{53.48} & \textbf{77.61} \quad 51.87 &  75.90 \quad 53.23\\
& 2 & 68.18 \quad 51.83 & 76.68 \quad 52.72 & 76.68 \quad 54.83 & 70.33 \quad 52.85 &  \textbf{77.15} \quad \textbf{55.75}\\
& 4 & 74.30 \quad 51.97 & 79.58 \quad 50.96 & 76.02 \quad \textbf{53.97} & 74.22 \quad 52.19 &  \textbf{79.18} \quad 51.71\\
& 6 & 53.90 \quad 48.43 & 57.78 \quad 48.73 & 53.90 \quad \textbf{51.43} & 51.80 \quad 48.55 &  \textbf{60.21} \quad 50.03\\
& Avg & 66.28 \quad 50.98 & 72.16 \quad 51.17 & 66.28 \quad 50.98 & 65.99 \quad 51.37 &  \textbf{73.11} \quad \textbf{52.68}\\
\hline
\end{tabularx}
\caption{Baseline Model Performance \& Hybrid Model Performance for AUC and pAUC Metrics. Highest means good.\emph{ Note: Best value achieved at each Machine ID denoted in Bold}}
\label{tab:performance-comparsion-hybrid}
\vspace{-1.5em}
\end{table*}
}

\subsection{Evaluation Results}
For evaluation purposes, this study incorporated Area under the Receiver Operating Characteristic (ROC) Curve (AUC) and Partial AUC(pAUC). The pAUC(Equation \ref{eq:pAUC}) is a notable metric generated with an AUC calculated from a portion of the ROC curve over the pre-specified range of interest. pAUC is calculated as the AUC over a low false-positive-rate (FPR) range [0,p], where p is the explicitly defined threshold. As many related studies did, This work set p to $5\%$ as many related studies incorporate p as $5\%$, which makes it easy to compare with the related studies.

\begin{equation} 
\label{eq:pAUC}
pAUC = \frac{1}{[pN^-] N^+}\sum_{i=1}^{[pN^-]}\sum_{j=1}^{N^+}H(A_\theta(x_j^+) - A_\theta(x_i^-))   
\end{equation}

The performance assessment of the hybrid architecture is displayed in an informative and illustrative manner in the table referred to as Table \ref{tab:performance-comparsion-hybrid}. It is clear from the results table that there is not a single model deemed superior to all of the machine types. For example, CVAE model performed exceptionally well on the Slider machine but fared poorly on the Valve and Fan machine. However, as a whole, the TCN-CVAE model has a better AUC, pAUC value on every machine on average, and it was able to reach the best value (in bold) on 70\% of the machine IDs. In order to compare these results we just consider Avg AUC and pAUC values obtained on each machine after this point.

\subsection{Compare with the Related Studies}

We consider the average values obtained from our study (with the presented data in Table \ref{tab:performance-comparsion-hybrid}) to compare with the related studies and depicted in Table \ref{tab:state-of-the-art-comparsion}.

{
\renewcommand{\arraystretch}{0.9}
\vspace{-1.0em}
\begin{center}
\begin{table*}[t]
\centering
\begin{tabularx}{\linewidth}{>{\bfseries}c@{\hspace{1cm}} c@{\hspace{1cm}} c c@{\hspace{1cm}} c@{\hspace{1cm}} c@{\hspace{1cm}} c@{\hspace{1cm}} c}
\hline
 Study & & Our & DCASE2020 & \cite{Jalali2020} & \cite{Ribeiro2020} & \cite{Giri2020} & \cite{daniluk2020}\\
\hline
Model & & TCN-CVAE & Dense AE & LSTM-AE & CAE & MobileNetV2 & Ensemble\\
\hline
Param & & 1,764,449 & 269,992 & 755,776 & 4,000,000 & 73,450,548 & 60,941,952\\
\hline
\multirow{2}{*}{Fan} & AUC & 70.29 & 65.85 & 67.32 & 66.78 & 82.33 & 94.12 \\
               & pAUC & 52.81 & 52.45 & 52.05 & 52.63 & 78.97 & 88.23 \\
\hline
\multirow{2}{*}{Pump} & AUC & 75.06 & 72.89 & 73.94 & 72.07 & 86.94 & 97.31 \\
                & pAUC & 62.58 & 59.99 & 61.01 & 60.96 & 79.60 & 92.56\\
\hline
\multirow{2}{*}{Slider} & AUC & 87.73 & 84.76 & 84.99 & 91.77 & 97.28 & 97.85 \\
                & pAUC & 67.78 & 66.53 & 67.47 & 76.20 & 89.54 & 94.54\\
\hline
\multirow{2}{*}{Valve} & AUC & 73.11 & 66.28 & 67.82 & 78.83 & 97.38 & 98.35 \\
                & pAUC & 52.68 & 50.98 & 51.07 & 53.10 & 91.21 & 92.11\\
\hline
\end{tabularx}
\caption{Hybrid Model Performance for AUC and pAUC Metrics compare to related studies. The highest means good.\emph{ Note: Best value achieved at each Machine ID denoted in Bold}}
\label{tab:state-of-the-art-comparsion}
\vspace{-1.5em}
\end{table*}
\end{center}
}

It simplified the process of conveying the effectiveness of the proposed TCN-CVAE architecture and positioning it with the state-of-the-art. At first, it is confirmed that our model is not state-of-the-art since there is a clear gap in AUC, pAUC values compared to the DCASE2020 challenge winner's models of \cite{Giri2020} and \cite{daniluk2020}. But compared to the size of parameters, our model is lightweight and still produces elegant results on each machine type even outperforming the recurrent neural network architectures of \cite{Jalali2020}. Another critical point is that the winning models did not follow the reconstruction-based technique while they individually changed the frame size and window size for the different machines, while our technique follows the same window size and framing for all the machine types. Moreover, they used computer vision-based improvement like data augmentation. Most importantly, none of the studies have mentioned how well they learn representation in a visualized form. But in our work, we present what those numbers actually mean and how the learning is actually stated in the context of data space in \textbf{Representation Improvment} experiment.

Moreover, this finding led us to maintain our prior hypothesis, which states that mono architectures do not perform better when considered as a whole and do not function better when considered individually. Both of these models have benefits and drawbacks. The hybrid model technique promotes improved feature learning and more accurately infers anomalies since it combines aspects of each model, which leads to an improvement of AUC and pAUC values while reducing the reconstruction better than single architecture. Further representation improvement states that detecting anomalies is easy for the model when the latent space has a suitable mechanism for standardizing via architectures like Variational Auto Encoders.

\subsection{Representation Improvement}

This experiment delivers a unique experiment to visually prove what the numbers really interpret in Table \ref{tab:performance-comparsion-hybrid} as our ultimate goal of learning the normal behaviour of MIMII machine types. We strive to prove hybrid architectures improve the concept of normality via representation learning with the support of t-SNE visualizations.

\begin{center}
\begin{figure}[htbp]
\vspace{-1.5em}
\centerline{\includegraphics[scale=0.40]{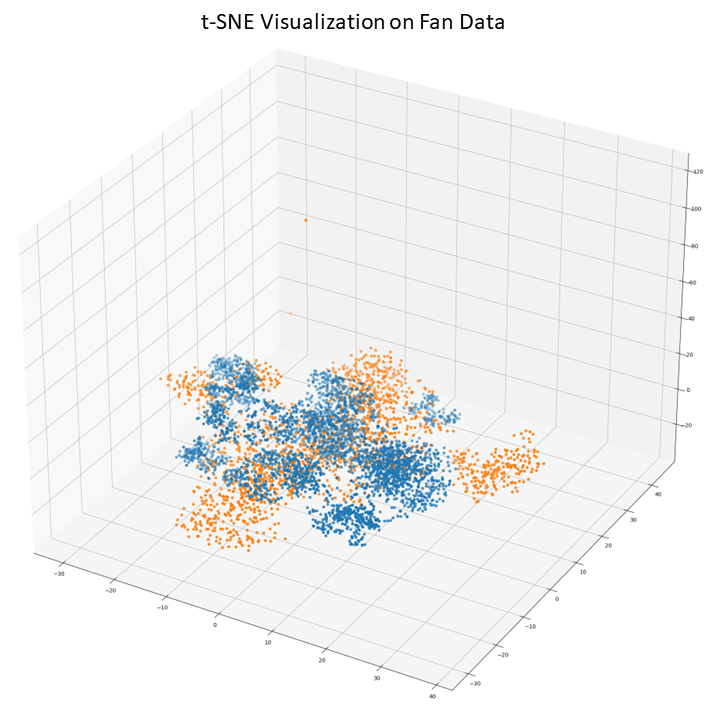}}
\caption{3D plot of Fan data in t-SNE visualization after applying TCN-CVAE modelling. Observe the representation improvement with \ref{fig:tSNEmimii-3d-fan}. Blue data items are normal data instances and orange data items are anomaly data instances}
\label{fig:tSNEfan3d}
\vspace{-1.5em}
\end{figure}
\end{center}

Also, it follows in light of the findings from Experiment 1, the continuation(2nd stage of Experiment 1) of tests was carried out with the goal of \textbf{graphically representing the complete data distribution after training the hybrid model}. This was done with the intention of gaining an overall grasp of the circumstance by gradually expanding the dimensions. You may view the 3D plot of the t-SNE representation of how the data from the fan machine is presented by looking at the figure that is referred to as \ref{fig:tSNEfan3d}. At this stage, the dimensionality is reduced from 5632 to 3, meaning that data with a high dimension (raw audio after Mel-Spectrogram feature extraction) is reduced by three dimensions after the encoding process. The fact that the normal (shown in blue) and the abnormality (shown in orange) zones still overlap areas exemplifies how challenging it could be to identify anomalies in the high-level stage too. 

However, on the bright side, it was able to convey an essential point by stretching anomalous samples throughout the regular data area. The above representation delivers a key point, which is that, when the anomalous (orange) data instance wraps around the normal data instances in a 2-D plot where there is overlap, this does not indicate that it is not precisely in every direction. What this indicates is that differentiation is achievable even in high-dimensional data spaces which is why the TCN-CVAE model results in a lower anomaly score than the normal instance while the anomaly instance result in a much greater anomaly score. Examine the transition from the X-Y to the Z axis in the following Figure \ref{fig:xyz-axis-3d-plot} with great care. It is possible to perceive that the blue sample is covered up by orange dots along each axis like \textbf{icing of a cake}. As a result, we may draw the conclusion that \textbf{anomalies exhibit one-of-a-kind patterns in higher dimensions and thus our modelling boundary can be differentiated as a space in dimensions}. Also much more importantly our initial hypothesis is depicted in Figure \ref{fig:anomaly-cloud} Part 3 is achieved to a certain level with our proposed hybrid architecture. Why it is so, still there are some orange instances inside that bounding box means, there is still room for understanding the exact normal data space.

\vspace{-0.5em}
\begin{center}
\begin{figure}[htbp]
\centerline{\includegraphics[scale=0.30]{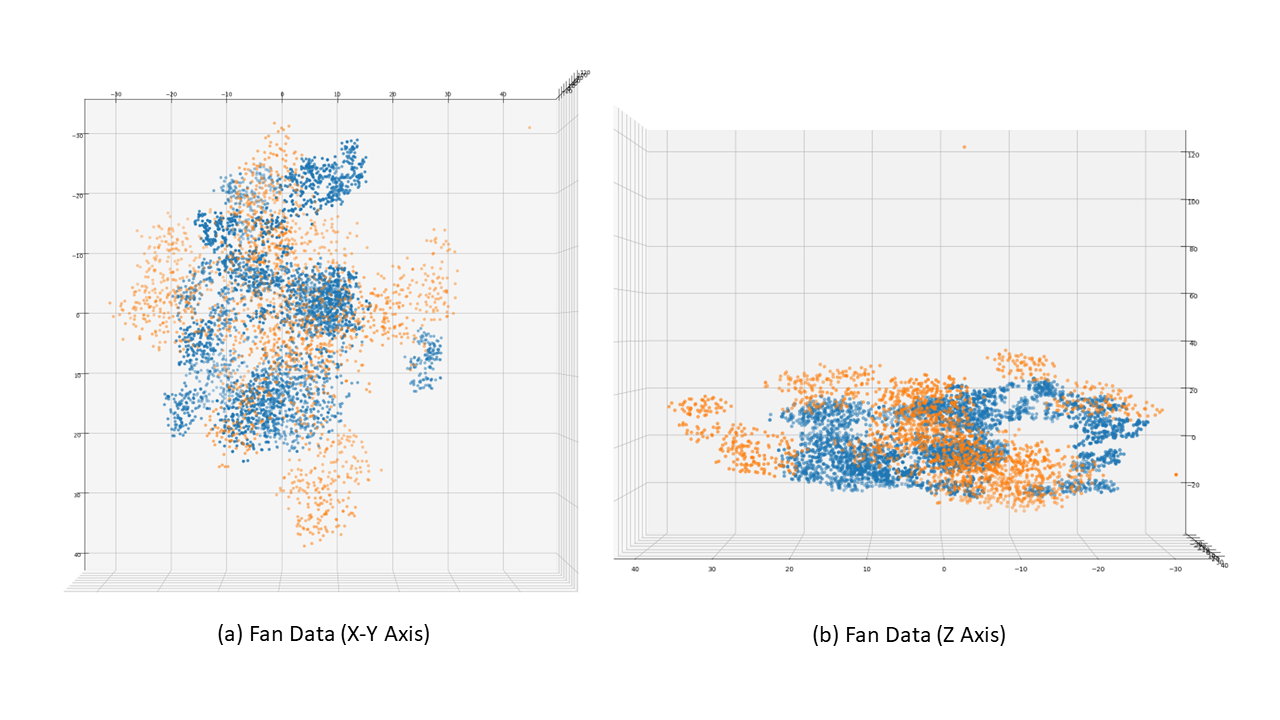}}
\caption{3D plot of Fan data in t-SNE visualization in X-Y and Y-Z Axis to determine the data distribution in axis. Blue data items are normal data instances and orange data items are anomaly data instances}
\label{fig:xyz-axis-3d-plot}
\vspace{-2.2em}
\end{figure}
\end{center}

\section{Discussion \& Conclusion}
Detecting anomalous behaviours helps to find new knowledge of a given phenomenon since anomaly behaviours rarely occur and give better insights to prevent those instances in the future. Usually, anomaly behaviours happen unexpectedly due to the unconscious nature of the objects. This study will directly or as a first insight helps to prevent happening severe anomaly behaviours generated in real-time via helping to diagnose them as early as possible.

The identification of the temporal convolutional-based TCN architecture with a Variational Auto Encoder as a sophisticated hybrid model for detecting acoustic anomalies marks a significant advancement. The proposed hybrid model architecture demonstrates substantial improvements in results compared to individual model performances, as corroborated by the data presented in Table \ref{tab:performance-comparsion-hybrid} and Table \ref{tab:state-of-the-art-comparsion}. These findings affirm the effectiveness of our initial hypothesis advocating for hybrid modelling with a blend of different strategies in addressing complex anomalies within the acoustic machine condition domain. Positioning our model alongside state-of-the-art alternatives reveals its lightweight nature, boasting 50 times fewer parameters than \cite{Giri2020} \cite{daniluk2020}. Noteworthy attributes include speed of convergence and lower training time compared to other complex architectures. At the same time, it delivers promising results by passing over AUC and pAUC values in the related single-model architectures present in the literature. The amalgamation of these characteristics establishes the proposed model as well-suited and feasible for RTAAD problems, meeting the mandatory requirements for an effective real-time anomaly detection system.

As the proposed architecture is based on a semi-supervised reconstruction-based representation learning technique, it allows a distinct advantage over other learning approaches, visualizing the current distribution of the data and trying to define the boundary that defines the normal and bound to it. Throughout the experiments, sophisticated methods of defining this boundary were explored, revealing critical insights; when dimensionality increases normal boundary is easily distinguishable and when the number of model strategies improved representation ability. The last experiment findings further support those claims. However, this study not proposing a global model for RTAAD for MCM. So, it has a limitation of machine-dependent nature. But further, researchers are focusing on One-Shot-Learning to mitigate this limitation as the next step. 

\section{Future Directions}
Enhancing the proposed study flows in several strategic directions. Firstly, there is a need to systematically compare the performance of the suggested semi-supervised reconstruction-based hybrid models across diverse contexts, including both non-domain shifted and domain-shifted environments. This comparative analysis will contribute to a comprehensive understanding of the model's adaptability.

Additionally, the integration of cutting-edge sequence models, such as the Transformer, into the framework is imperative for advancing the capabilities of acoustic anomaly detection. This inclusion will augment the study's technological sophistication and align it with the current state-of-the-art in the field.
Moreover, an exploration into the feasibility of real-time application is warranted to assess the practical implications of the proposed model. This extension will shed light on the model's responsiveness and utility in dynamic, real-world scenarios. Future investigations could dive into the implementation of an ensemble mechanism, employing averaging or other mathematical techniques, for feature extraction methods.

Recognizing the potential of Gammantone Filters \cite{spafe} as an alternative candidate for feature extraction, in conjunction with the Log-Mel features utilized in this study, opens avenues for more intricate research studies. Complex inquiries under the umbrella of representation learning techniques could be pursued, delving into questions such as the equivalence between the end of a normal boundary and the commencement of an anomalous boundary. "Is the end of boundary normal equivalent to the starting point of boundary anomaly?", "How can we dynamically change the representation for the normal boundary with domain-shifted natures?", and "Upto what extend normal sample should bottleneck(percentage-wise) or degrade in order to find the unique natures of anomalies?" present promising research avenues based on the insights derived from this study.

\bibliographystyle{ACM-Reference-Format}
\bibliography{icmlc-22}

\end{document}